\documentclass{trbunofficial}

\begin{document}


\title{SimTIO: A Simulation-Grounded Multi-Agent LLM Framework for Compositional Traffic Intervention Optimization}

\TRBauthor{Shuyang Li}{Graduate Student\\Civil and Environmental Engineering, Rensselaer Polytechnic Institute}{lis36@rpi.edu}[Troy, New York, 12180][0009-0006-8588-2947]
\TRBauthor{Ruimin Ke}{Assistant Professor\\Civil and Environmental Engineering, Rensselaer Polytechnic Institute}{ker@rpi.edu}[Troy, New York, 12180]

\AuthorHeaders{Shuyang Li, Ruimin Ke}

\maketitle

\section{Abstract}
\noindent\textbf{Objectives:}~Traffic analysts must translate diagnosed
bottlenecks into executable interventions while avoiding local improvements
that degrade network-wide performance. This study develops SimTIO, a
simulation-grounded multi-agent large language model framework for generating,
composing, and selecting traffic interventions under explicit operational
constraints.

\hfill\break%
\noindent\textbf{Methods:}~An unmodified SUMO scenario was simulated to derive
vehicle- and edge-level traffic measures and a baseline-frozen set of ten
bottleneck edges. A grounded sampler initialized signal-control,
corridor-speed, and demand-preserving routing interventions, and three
specialist agents used measured simulation feedback to select one-parameter
refinements from validator-confirmed mutation catalogs. Each case used at most
seven candidate simulations: three initial actions, up to two feedback
mutations, and up to two compatible two-action bundles. The evaluation covered
five U.S. urban networks, three synthetic-demand seeds, and 2,400
origin--destination trips per scenario; final selection bounded global delay,
neighboring-road spillover, throughput, and teleport events.

\hfill\break%
\noindent\textbf{Findings:}~Across 15 city--seed cases, the LLM-feedback
optimizer reduced Top-10 bottleneck time loss by
$9.18\%\pm10.18\%$ and network-wide delay by 2.78\%. It found a feasible
improving plan in 86.7\% of cases, compared with 73.3\% for grounded random
search and 80.0\% for a deterministic heuristic under the same simulation
budget. Only three of the 13 applied LLM plans were unmodified initial actions;
six were feedback-refined singles and four were bundles. Paired Wilcoxon tests
found no statistically significant Top-10 advantage over either baseline
($p=0.407$ and $p=0.639$).

\hfill\break%
\noindent\textbf{Novelty:}~SimTIO separates LLM-guided feedback refinement from
simulation-based decision authority and evaluates heterogeneous intervention
compositions rather than relying on predicted or additive effects. Its
baseline-frozen objective and no-operation guard prevent unsafe local gains
from determining the final plan.

\hfill\break%
\noindent\textbf{Practical Applications:}~The framework can support traffic
analysts in screening signal, speed-limit, and routing interventions before
field deployment. Practical use requires locally calibrated demand, verified
network data, and agency review of the simulation-selected intervention.

\newpage

\section{Introduction}
\label{sec:intro}

Microscopic traffic simulation is widely used to evaluate congestion, signal
control, routing, emissions, and network-level mobility.  Simulators such as
SUMO provide detailed measurements and allow interventions to be tested without
affecting a real transportation system.  However, simulation is primarily an
evaluation tool: converting a diagnosed bottleneck into an executable action
still requires an analyst to select a control strategy, modify interdependent
simulation artifacts, and evaluate possible network-wide side effects.  This
gap between diagnosis and action limits the automation of simulation-based
traffic optimization.

Large language models (LLMs) offer a flexible way to reason across heterogeneous
interventions, including signal retiming, corridor control, and demand
rerouting.  Nevertheless, plausible language is not evidence of traffic
benefit.  An LLM may reference nonexistent network entities, request unsupported
operations, omit required parameters, or propose an intervention that improves
a local bottleneck by transferring congestion elsewhere.  Traffic actions may
also interact: two individually reasonable interventions can reinforce or
interfere with one another.  Because exhaustive simulation of all combinations
is impractical, an LLM-based optimizer must ground proposals in executable
capabilities, search combinations within a limited budget, and reserve final
selection for measured simulation outcomes.

This study develops a simulation-grounded multi-agent LLM traffic optimizer.
The framework first simulates the unmodified scenario and extracts compact
bottleneck-centered evidence containing traffic measurements, local
connectivity, and relevant signal identifiers.  A grounded sampler initializes
one validator-confirmed action from each available signal-control,
corridor-speed, and demand-routing family.  After those actions are simulated,
three specialist agents use the measured outcomes to select local refinements
from program-generated legal mutation catalogs.  A deterministic capability
registry rejects unsupported actions, invalid identifiers, and incomplete
parameters before execution.  The LLM therefore serves as a feedback-guided
local search operator rather than as the final traffic-control decision maker.

The system further performs a simulation-feedback beam search.  Measured KPIs
and constraint violations are returned to the responsible specialist, which
produces an auditable one-parameter refinement of a retained parent.  The
remaining budget is used for pairwise search over compatible interventions; if
no unseen legal refinement or compatible bundle exists, that portion of the
budget remains unused.
Candidate actions are applied sequentially and re-simulated, so their combined
effect is measured rather than inferred from textual predictions.  Final
selection minimizes time loss on ten baseline-frozen bottleneck edges while
allowing at most a 1\% network-wide delay increase and 5\% neighboring-road
spillover, while prohibiting throughput loss or additional teleport events.
The unmodified scenario remains available as a no-operation guard.  An offline
Pareto audit additionally checks whether the delay-selected plan is dominated
on measured delay, emissions, or mobility outcomes.

The primary evaluation covers five urban networks, three demand seeds, and
2,400 origin--destination trips per scenario.  Across 15 runs, the selected
outcomes reduce Top-10 bottleneck time loss by $9.18\%\pm10.18\%$ and
network-wide delay by 2.78\%.  The LLM-feedback search finds a feasible
improving plan in 86.7\% of cases, compared with 73.3\% for grounded random
search and 80.0\% for a deterministic heuristic under the same simulation cap.
Only three of the 13 applied LLM plans are unmodified initial actions; six are
feedback-refined singles and four are bundles.

This work contributes a simulation-grounded multi-agent framework that generates
heterogeneous traffic interventions from compact diagnostic and topological
evidence while validating every proposal against executable capabilities.  It
introduces a budgeted composition mechanism that applies and re-simulates
compatible intervention pairs, allowing measured interactions to influence
selection without exhaustive search.  The framework further combines a
baseline-frozen bottleneck objective with constraints on global delay,
spillover, throughput, and simulation integrity, together with a no-op guard
and an offline Pareto dominance audit.  Finally, the multi-network, multi-seed
evaluation quantifies both achievable improvements and the rejection of locally
promising but globally unsafe candidates.  Together, these components position
the specialist LLMs as role-specific, feedback-guided local search operators
while reserving final authority for executable tools, simulation evidence, and
explicit optimization constraints.

\section{LITERATURE REVIEW}

\subsection{Execution-Grounded LLM Algorithm Design}

Large language models have increasingly moved from one-shot solution
generation toward feedback-driven optimization. OPRO treats an LLM as a
black-box optimizer by conditioning new proposals on previously evaluated
solution--score pairs~\cite{yang2024opro}, while Self-Refine demonstrates how
verbal feedback can support iterative revision without additional
training~\cite{madaan2023selfrefine}. For executable algorithm discovery,
FunSearch couples an LLM with deterministic evaluation and an evolutionary
program database~\cite{romera2024funsearch}. Evolution of Heuristics evolves
both heuristic ideas and executable implementations~\cite{liu2024eoh}, and
ReEvo converts performance comparisons into reflective feedback for subsequent
generations~\cite{ye2024reevo}. MCTS-AHD organizes candidates in a search tree
to improve exploration~\cite{zheng2025mctsahd}, whereas AlphaEvolve scales the
same generate--evaluate--retain principle through an evolutionary coding-agent
pipeline with automated evaluators~\cite{novikov2025alphaevolve}. Collectively,
these studies show that LLM proposals become more reliable when validity and
quality are determined by external execution rather than linguistic judgment
alone. Our framework adopts this principle, but its search objects are
structured traffic interventions and its evaluator is a microscopic traffic
simulator.

\subsection{Multi-objective, Compositional, and Multi-agent Search}

Most LLM-based algorithm-design systems search for a single program or
heuristic under a scalar score. MEoH extends this setting by maintaining
non-dominated heuristics and accounting for both objective-space dominance and
program diversity~\cite{yao2025meoh}. E2OC addresses interdependent operator
design by representing composition as a sequential decision process and using
Monte Carlo tree search~\cite{qiu2026e2oc}, while OptiMUS illustrates a
complementary form of execution grounding in which LLM-generated optimization
models are checked and solved by mathematical-programming
solvers~\cite{ahmaditeshnizi2024optimus}. Multi-agent debate further suggests that
role separation and critique can improve proposal diversity and reasoning
quality~\cite{du2024debate}. In our framework, signal, network, and routing
agents therefore select feedback-guided refinements for complementary
intervention families, but neither the agents nor their predicted effects
determine the final plan. Valid single interventions are evaluated independently,
promising candidates are combined subject to
compatibility and evaluation-budget constraints, and the baseline, singles,
and bundles form a common candidate pool. A constraint-aware selector is used
for the primary experiment, while an offline Pareto dominance audit checks
secondary traffic outcomes without changing the selected plan, following the
general non-dominated-selection principle used in multi-objective
optimization~\cite{deb2002nsgaii}.

\subsection{LLMs for Traffic Control and Simulation-Grounded Optimization}

Before LLM-based control, learning-based traffic-signal research developed
simulator-trained policies such as the max-pressure-inspired
PressLight~\cite{wei2019presslight} and the graph-attention-based
CoLight~\cite{wei2019colight}. Recent LLM approaches instead use language models for
traffic reasoning and decision support. LLM-Assisted Light augments an LLM with
traffic perception and control tools~\cite{wang2024llmassistedlight}; LLMLight
directly selects signal phases from structured
observations~\cite{lai2025llmlight}; and CoLLMLight extends this paradigm to cooperative,
network-wide control~\cite{yuan2026collmlight}. Hybrid methods such as iLLM-TSC
inspect and correct reinforcement-learning actions under abnormal or partially
observed conditions~\cite{pang2024illmtsc}, whereas EvolveSignal searches for
executable signal-control programs using simulator
feedback~\cite{wang2025evolvesignal}. A related line of work uses LLM agents to automate
traffic-simulation workflows: ChatSUMO translates natural-language requests
into SUMO scenarios~\cite{li2024chatsumo}, ChatSUMO-Agent adds closed-loop
planning, tool execution, and refinement~\cite{li2026chatsumoagent}, and the
concurrent TrafficSimAgent preprint studies hierarchical agents for general
traffic-simulation tasks~\cite{du2025trafficsimagent}. In contrast, our focus is
not scenario generation or stepwise signal control, but simulation-grounded
search over heterogeneous network, routing, and signal interventions. SUMO
provides the executable microscopic evaluator~\cite{lopez2018sumo}, while the
cross-layer formulation is motivated by the established coupling between
signal control and route assignment~\cite{meneguzzer1997combined} and between
network design and travelers' route choices~\cite{yang1998networkdesign}.


\section{Methodology}
\label{sec:methodology}

\subsection{Framework and Traffic Evidence}
\label{sec:framework-evidence}

\begin{figure}[htbp]
    \centering
    \includegraphics[width=\linewidth]{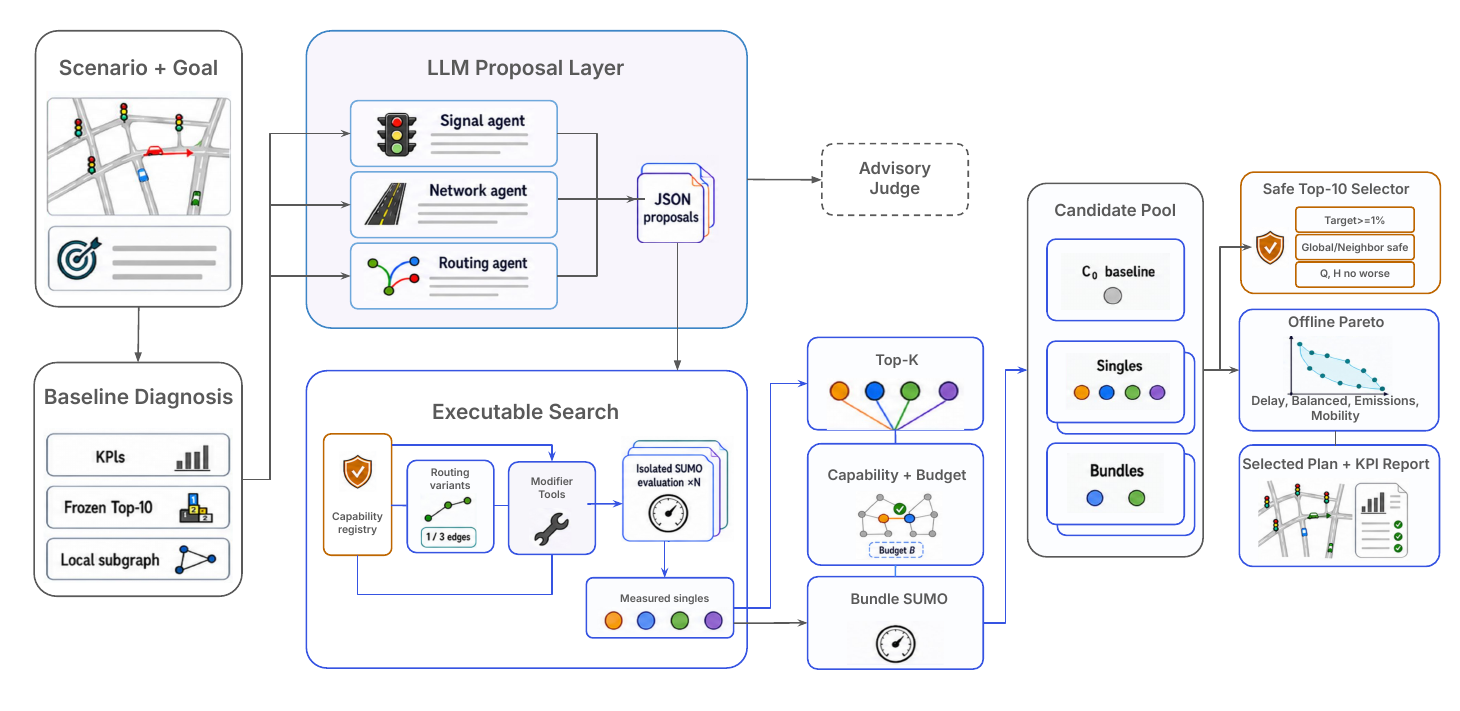}
    \caption{Simulation-grounded multi-agent traffic optimization framework.
    Grounded sampling initializes diverse interventions; specialist LLM agents
    refine measured parents, after which compatible actions may be composed
    before constrained final selection.}
    \label{fig:framework}
\end{figure}

The proposed framework separates grounded exploration, LLM-guided refinement,
and simulation-based decision making.  As shown in Figure~\ref{fig:framework}, an
unmodified SUMO scenario is first simulated to establish a baseline.  A
deterministic diagnosis module extracts traffic measurements and a compact
bottleneck-centered subgraph.  A seeded grounded sampler initializes one
intervention from each control family, which is validated and evaluated
independently.  A
feedback-guided beam then sends measured outcomes back to the responsible
specialists for one LLM-selected local refinement.  The remaining budget is
used for compatible two-action bundles.  If no unseen legal refinement or
compatible bundle is available, the budget remains unused rather than being
filled by a deterministic backfill policy.  The best measured candidate is
selected subject to explicit safety constraints.  The LLM supplies grounded
search directions, whereas the orchestrator enumerates legal local alternatives
and SUMO provides the decision evidence.

The input consists of a SUMO network, an OD-demand file, and a simulation
configuration.  The baseline run produces vehicle-level \texttt{tripinfo} and
60-second edge measurements.  The diagnosis aggregates speed, density,
occupancy, waiting time, time loss, and traffic exposure.  For each edge $e$, it
computes

\begin{equation}
\begin{split}
C_e ={}& 0.35(1-r_e)
       +0.25\,\phi(\rho_e;35)
       +0.20\,\phi(o_e;30) \\
     & +0.10\,\phi(L_e;300)
       +0.10\,\phi(W_e;300),
\end{split}
\label{eq:congestion-score}
\end{equation}

where $r_e$ is the observed-to-permitted speed ratio, $\rho_e$ is density,
$o_e$ is occupancy, $L_e$ and $W_e$ are accumulated time loss and waiting time,
and $\phi(x;h)=\min(1,\max(0,x/h))$.  The score $C_e$ is computed once from the
unmodified baseline and is used only as a diagnostic ranking index.  It is not
the objective optimized by SimTIO and is not recomputed after an intervention.
To reduce sensitivity to microscopic network segmentation, non-internal edges
shorter than 10 m are excluded before ranking.  Let
$\mathcal{E}_{\mathrm{eligible}}$ denote the remaining edges; the fixed target
set is

\begin{equation}
    \mathcal{T}
    =\operatorname{Top10}_{e\in\mathcal{E}_{\mathrm{eligible}}}(C_e).
    \label{eq:baseline-target-set}
\end{equation}

Thus, $C_e$ determines where the optimizer searches, whereas candidate quality
is determined by the simulated total time loss $L_{\mathcal{T}}$ on the fixed
set $\mathcal{T}$.  Up to five highest-ranked edges also seed a one-hop local
subgraph, limited to 80 edges, that provides agents with valid neighbor,
junction, and traffic-light identifiers without inserting the full network XML
into their prompts.

Freezing $\mathcal{T}$ prevents a moving target in which an intervention appears
effective merely because congestion is shifted to different edges.  The local
subgraph retains enough topology to identify contiguous corridors and nearby
controllers while limiting prompt length.

\subsection{Optimization Problem Formulation}
\label{sec:optimization-formulation}

The optimization object is an executable intervention plan, not an LLM's prose
recommendation or predicted KPI value.  We represent a plan as

\begin{equation}
\mathbf{x}=
\left(x^{\mathrm{sig}},x^{\mathrm{net}},x^{\mathrm{route}}\right)
\in\mathcal{X}_{\mathrm{cap}},
\label{eq:decision-vector}
\end{equation}

where an inactive component is denoted by $\emptyset$.
$x^{\mathrm{sig}}$ specifies a traffic-light controller and a fixed-time phase
plan derived using Webster's signal-timing method~\cite{webster1958traffic};
$x^{\mathrm{net}}$ specifies either a corridor-speed
change, an edge removal, or a one-way conversion together with its target and
discrete parameter; and $x^{\mathrm{route}}$ specifies an avoided edge and the
rerouted fraction.  The capability-grounded domain
$\mathcal{X}_{\mathrm{cap}}$ contains only schema-valid actions, scenario-valid
identifiers, and supported discrete values.  A single plan activates one
component, while a compatible bundle activates at most two components.  This
cardinality limit and the compatibility rules prevent the optimizer from
combining conflicting edits or changing the same control family twice.

Let the deterministic SUMO evaluation map be

\begin{equation}
\mathcal{S}(\mathbf{x})=
\left(L_{\mathcal{T}},L_{\mathcal{N}},D,Q,H,E,V\right),
\label{eq:simulator-response}
\end{equation}

where $L_{\mathcal{T}}$ is total time loss on the baseline-frozen Top-10 edges,
$L_{\mathcal{N}}$ is time loss on their one-hop neighbors, $D$ is network-wide
mean vehicle time loss, $Q$ is throughput, $H$ is the teleport count, $E$ is
total $\mathrm{CO}_2$, and $V$ is mean speed.  Superscript $0$ below denotes
the response of the unmodified scenario $\mathbf{x}_0$.  We also define
$A(\mathbf{x})=1$ when all demand records and edge-level OD pairs are
preserved.  Let $\mathcal{C}_{\mathrm{eval}}\subseteq
\mathcal{X}_{\mathrm{cap}}$ denote the plans actually evaluated within the
simulation budget.  The optimizer returns the best measured feasible incumbent:

\begin{equation}
\begin{aligned}
\mathcal{F}_{\mathrm{eval}}=\{\mathbf{x}\in\mathcal{C}_{\mathrm{eval}}:\quad
&A(\mathbf{x})=1,\quad
L_{\mathcal{T}}(\mathbf{x})\leq0.99L_{\mathcal{T}}^0,\\
&D(\mathbf{x})\leq1.01D^0,\quad
L_{\mathcal{N}}(\mathbf{x})\leq1.05L_{\mathcal{N}}^0,\\
&Q(\mathbf{x})\geq Q^0,\quad H(\mathbf{x})\leq H^0\},\\
\hat{\mathbf{x}}&=
\begin{cases}
\arg\min_{\mathbf{x}\in\mathcal{F}_{\mathrm{eval}}}L_{\mathcal{T}}(\mathbf{x}),
&\mathcal{F}_{\mathrm{eval}}\neq\emptyset,\\
\mathbf{x}_0,&\mathcal{F}_{\mathrm{eval}}=\emptyset.
\end{cases}
\end{aligned}
\label{eq:top10-constrained-selection}
\end{equation}

Thus, the optimized quantity is frozen-bottleneck time loss.  The 1\% Top-10
requirement is an acceptance threshold; global delay and neighboring time loss
are bounded spillover constraints; and demand integrity, throughput, and
teleports are operational constraints.  Emissions and speed are reported as
secondary outcomes and enter only the offline Pareto dominance audit.  This
separation prevents an LLM-predicted effect from being mistaken for the
objective used by the primary optimizer.  Because
$\mathcal{C}_{\mathrm{eval}}$ is budget limited, $\hat{\mathbf{x}}$ is the best
evaluated incumbent rather than a claim of global optimality over
$\mathcal{X}_{\mathrm{cap}}$.

\subsection{Grounded Random Initialization}
\label{sec:grounded-proposals}

The diagnosis defines a discrete search space over the baseline-frozen
objective edges.  Before sampling, the executable capability validator removes
non-optimizable controllers, missing reverse-edge pairs, detectable local
topology violations, and malformed controls.  This pre-execution check confirms
that a candidate is supported and locally well formed; complete demand and OD
preservation are checked again after the modification is materialized.  Using
the recorded case seed, the primary optimizer samples one validator-confirmed
proposal from each available specialist family in round-robin order.  The
signal family selects an evidenced controller for fixed-time green-split
optimization based on Webster's method while preserving phase states and
yellow/all-red transition intervals.  The
network family samples a legal corridor-speed, edge-removal, or one-way action;
corridor speed factors belong to $\{0.7,0.8,0.9\}$.  The initial routing family
uses a one-edge trust region and chooses

\begin{equation}
    \alpha\in\{0.25,0.50,0.75,1.00\}
\end{equation}

as the rerouted fraction.  The first three samples are exactly the prefix of
the grounded-random baseline generated with the same seed.  Thus, the
LLM-feedback and grounded-random conditions share the same first three actions
and differ only in how the remaining simulation budget is allocated.

Only vehicles whose
reference routes traverse a selected edge are eligible, and avoidance is soft
so that endpoints and unavoidable connections remain valid.  Unselected
vehicles retain their reference routes, and every modified demand file must
preserve all records and OD pairs.

\begin{figure}[htbp]
    \centering
    \includegraphics[width=\linewidth]{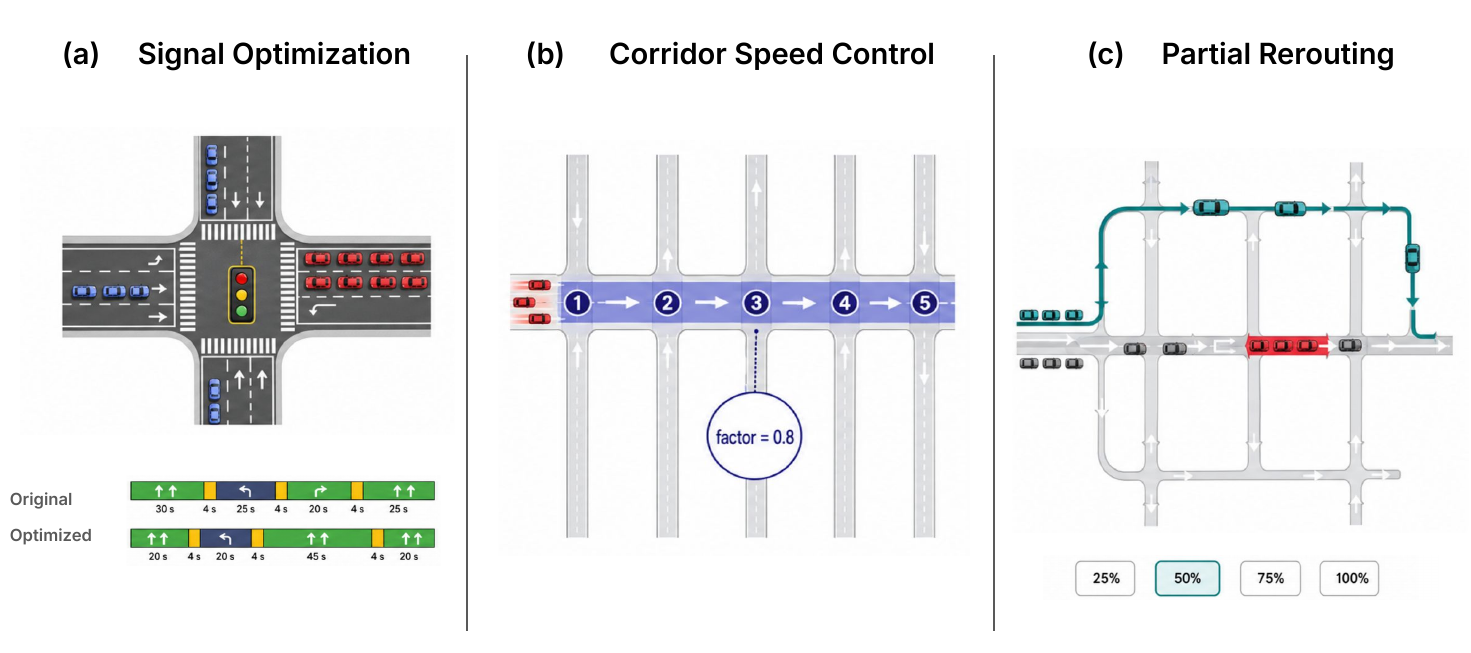}
    \caption{Intervention mechanisms supported by the optimizer: fixed-time
    signal green-split adjustment, corridor-level speed control, and partial
    rerouting around a diagnosed bottleneck.  Grounded initialization selects
    the initial targets and parameters, while an LLM feedback specialist selects
    a validator-confirmed one-parameter refinement.  Deterministic utilities
    apply the corresponding SUMO modifications.}
    \label{fig:intervention-mechanisms}
\end{figure}

Specialization prevents an agent from editing the entire scenario.  As
illustrated in Figure~\ref{fig:intervention-mechanisms}, signal proposals change
green allocation without altering phase states or transitions; corridor edges
are resolved deterministically from an evidenced anchor; and the routing
utility, rather than the LLM, selects eligible vehicles and verifies route
connectivity.  The model therefore supplies high-level search choices, while
deterministic code performs the network edits.

Each initialized action is encoded as a structured proposal containing an
executable action, target, parameters, and evidence references.  Before
execution, a deterministic capability registry checks that the action is
implemented, required artifacts and parameters are present, and every edge or
traffic-light identifier exists in the scenario.  Invalid fractions,
placeholders, unsupported operations, and incomplete proposals are rejected.

Validation is performed in two stages.  Schema validation first checks the
proposal type, required fields, numeric ranges, and argument names.  Scenario
validation then binds those arguments to the current SUMO files and checks
identifier existence and action-specific preconditions.  This design prevents
an unsupported proposal or mutation from entering the expensive simulation
stage and provides a trace from every executable modification back to baseline
evidence.

\subsection{Simulation-Grounded Candidate Search}
\label{sec:candidate-search}

Every validated random-initial proposal is applied to an isolated copy of the
baseline scenario and evaluated through a complete SUMO simulation.  All
candidates start from identical baseline artifacts, and a failed candidate does
not affect the remaining evaluations.  The resulting vehicle- and edge-level
outputs are processed using the same definitions as the baseline, producing
measured KPIs and explicit constraint violations for each intervention.

The feedback-guided configuration then retains a feasible-first beam
$\mathcal{B}_{t}$ of width $b$.  Ties are resolved by measured objective value,
while role diversity prevents minor variants from one intervention family from
occupying the entire beam.  For every parent $c\in\mathcal{B}_{t}$, its
specialist receives the measured KPI vector, Top-10 improvement, global and
neighbor changes, and failed constraints.  Deterministic code first enumerates
a finite catalog $\mathcal{M}(c)$ of unseen, validator-confirmed mutations that
retain the action and change exactly one decision argument.  The specialist
then selects one mutation identifier $m(c)\in\mathcal{M}(c)$ using the measured
feedback; it cannot introduce a new identifier or numeric value.  The update is

\begin{equation}
\mathcal{C}_{t+1}=\mathcal{C}_{t}\cup
\{m(c):c\in\mathcal{B}_{t}\},\qquad
\mathcal{B}_{t+1}=\operatorname{Top}_{b}^{\mathrm{feas,div}}
(\mathcal{C}_{t+1}),
\label{eq:feedback-beam-update}
\end{equation}

where every selected mutation is revalidated and re-simulated before it can
enter the next beam.  An invalid catalog selection receives one LLM repair
attempt.  Action changes, multi-argument rewrites, and previously tested
duplicates are rejected without consuming the SUMO budget.  The initial
configuration uses one grounded-initialization round followed by one
feedback-refinement round with beam width $b=2$; disabling the refinement round
recovers a one-shot search.  Candidate simulation
budget that cannot be used because a refinement is invalid or duplicated
remains unused; all feedback-guided mutations are chosen by the responsible LLM
specialist from the program-generated legal catalog.

After feedback refinement, the system retains the top $K$ single candidates and
enumerates compatible pairs within a remaining simulation budget $B$.  A bundle contains at most one action from each
intervention family, and actions that conflict on the same edge or controller
are rejected.  Compatible actions are applied in the order

\begin{equation}
\text{topology}\rightarrow\text{speed}\rightarrow
\text{routing}\rightarrow\text{signal},
\label{eq:bundle-order}
\end{equation}

so each action is rebound to the artifacts produced by the preceding step.
The complete bundle is then re-simulated rather than scored by adding its
constituents' predicted or isolated effects.  Successfully evaluated singles
and bundles form the final candidate pool $\mathcal{C}_{\mathrm{eval}}$.

The beam width, Top-$K$ filter, and bundle budget control the number of
additional SUMO runs.  Singles are ordered using their measured performance, after which the
compatibility checker excludes repeated intervention families and conflicting
targets, such as speed modification and soft avoidance of the same edge.  The
sum of single-candidate scores is used only to prioritize which compatible
pairs receive the limited simulation budget.  It is not treated as the bundle's
performance because two individually useful actions can reinforce, cancel, or
shift the effects of one another.  The bundle run therefore captures interaction
effects that cannot be inferred reliably from the isolated evaluations.  In
the reported experiments, $K=4$, at most two bundles are simulated, and the
total candidate budget is capped at seven SUMO evaluations per case; these
settings apply equally to the LLM and non-LLM proposal-source conditions.

\subsection{LLM Feedback Prompts}
\label{sec:feedback-prompts}

The three role-specific system prompts used during feedback refinement are
reported below. Grounded initialization is seeded and programmatic and therefore
has no LLM prompt. Each specialist receives the retained parent action, measured
SUMO outcomes, constraint violations, baseline evidence, and a program-generated
catalog of unseen validator-confirmed one-argument mutations. The model may
select one catalog identifier but cannot rewrite an action, invent an identifier,
or introduce a numeric value. Responses use JSON mode and are limited to 500
output tokens; an invalid selection receives one repair attempt. Bundle
construction, compatibility filtering, simulation, Pareto auditing, and final
selection remain deterministic.

\lstdefinestyle{llmprompt}{
    basicstyle=\ttfamily\scriptsize,
    frame=single,
    breaklines=true,
    breakatwhitespace=false,
    columns=fullflexible,
    keepspaces=true,
    showstringspaces=false,
    tabsize=2,
    aboveskip=0.7em,
    belowskip=0.7em
}

\subsubsection{Signal-Control Specialist}

\begin{lstlisting}[style=llmprompt]
You are the signal-control feedback specialist in a traffic optimization team.

The input contains:
- parent_proposal: the evaluated signal action;
- measured_outcomes: baseline-relative SUMO KPIs;
- constraint_violations: failed acceptance or safety constraints;
- legal_mutations: unseen validator-confirmed one-argument mutations.

Return EXACTLY ONE JSON object:
{
  "selected_mutation_id": "...",
  "reason": "...",
  "evidence_refs": ["..."],
  "risk": "..."
}

Rules:
- Select exactly one ID that appears in legal_mutations.
- Do not rewrite the action or arguments and do not invent an ID, TLS target,
  phase state, cycle length, split, or numeric value.
- Compare all catalog entries using the measured parent outcome and supplied
  baseline evidence; do not automatically choose the first entry.
- Prefer a mutation that addresses failed global-delay or spillover constraints
  while preserving the measured Top-10 benefit.
- Preserve yellow/all-red states and consider minor-movement starvation and
  downstream queue transfer.
- reason and risk must be substantive.
- If legal_mutations is empty, return selected_mutation_id="" and explain why.
\end{lstlisting}

\subsubsection{Road-Network Intervention Specialist}

\begin{lstlisting}[style=llmprompt]
You are the road-network feedback specialist in a traffic optimization team.

The input contains:
- parent_proposal: the evaluated network action;
- measured_outcomes: baseline-relative SUMO KPIs;
- constraint_violations: failed acceptance or safety constraints;
- legal_mutations: unseen validator-confirmed one-argument mutations.

Return EXACTLY ONE JSON object:
{
  "selected_mutation_id": "...",
  "reason": "...",
  "evidence_refs": ["..."],
  "risk": "..."
}

Rules:
- Select exactly one ID that appears in legal_mutations.
- Do not rewrite the action or arguments and do not invent an ID, edge target,
  corridor member, speed factor, or numeric value.
- Compare all catalog entries using the measured parent outcome and supplied
  baseline evidence; do not automatically choose the first entry.
- set_corridor_speed is a static corridor-level speed-limit adjustment; the
  deterministic tool resolves up to five same-road edges from one anchor.
- Prefer a mutation that repairs global-delay or neighbor-spillover violations
  without giving up the measured Top-10 benefit.
- For topology actions, consider OD disconnection, route loss, detour, and
  congestion transfer. For speed actions, consider queue smoothing and
  downstream spillback.
- reason and risk must be substantive.
- If legal_mutations is empty, return selected_mutation_id="" and explain why.
\end{lstlisting}

\subsubsection{Demand-Routing Specialist}

\begin{lstlisting}[style=llmprompt]
You are the demand-routing feedback specialist in a traffic optimization team.

The input contains:
- parent_proposal: the evaluated single-edge rerouting action;
- measured_outcomes: baseline-relative SUMO KPIs;
- constraint_violations: failed acceptance or safety constraints;
- legal_mutations: unseen validator-confirmed one-argument mutations.

Return EXACTLY ONE JSON object:
{
  "selected_mutation_id": "...",
  "reason": "...",
  "evidence_refs": ["..."],
  "risk": "..."
}

Rules:
- Select exactly one ID that appears in legal_mutations.
- Do not rewrite the action or arguments and do not invent an ID, avoided edge,
  rerouting fraction, or numeric value.
- The trust region contains exactly one avoided edge. Legal rerouting fractions
  are 0.25, 0.5, 0.75, and 1.0.
- Compare all catalog entries using the measured parent outcome and supplied
  baseline evidence; do not automatically choose the first entry.
- Prefer a mutation that reduces displaced congestion, excessive detour, or a
  failed global-delay or neighbor-spillover constraint while retaining local
  benefit.
- Every demand record and its exact edge-level OD must remain unchanged;
  rerouting is soft avoidance and may retain an unavoidable endpoint edge.
- reason and risk must be substantive.
- If legal_mutations is empty, return selected_mutation_id="" and explain why.
\end{lstlisting}

\subsection{Constrained Final Selection}
\label{sec:constrained-selection}

Demand preservation is a hard feasibility requirement.  A candidate is
discarded if it changes the number of demand records or any OD pair.  Throughput
is measured against the original source demand as
$Q=N_{\mathrm{completed}}/N_{\mathrm{source}}$, preventing an intervention from
appearing beneficial by silently removing difficult trips.

This formulation avoids assigning an arbitrary fixed delay penalty to
unfinished vehicles.  Instead, incomplete service is exposed directly through
throughput and teleport constraints.  Delay and time
loss are computed from the vehicles and edge measurements actually reported by
SUMO, while demand preservation ensures that the denominator cannot be changed
by an intervention.  The same aggregation code and simulation horizon are used
for the baseline and every candidate.

Equation~\ref{eq:top10-constrained-selection} is applied only to measured SUMO
outcomes.  An intervention must therefore produce at least a 1\% bottleneck
improvement, may increase global delay by at most 1\%, may increase neighboring
time loss by at most 5\%, and cannot reduce throughput or increase teleports.
The unmodified scenario $\mathbf{x}_0$ acts as a no-operation guard.  

The rule distinguishes target improvement from network-wide acceptability.
$L_{\mathcal{T}}$ tests whether the diagnosed bottlenecks improve,
$L_{\mathcal{N}}$ detects immediate spillover, and $D$ prevents a local gain
from being purchased through worse overall travel conditions.  Throughput and
teleports serve as operational safeguards.  The 1\% threshold suppresses
changes too small to be operationally meaningful, while
the neighbor tolerance permits limited redistribution around a bottleneck.  If
several candidates pass all requirements, minimizing frozen Top-10 time loss
keeps the final decision aligned with the stated local optimization objective.

\subsection{Pareto Dominance Audit}
\label{sec:pareto-method}

To check whether the delay-oriented selector discards an unambiguously better
candidate, an offline analysis examines each saved candidate pool using four
baseline-relative gains:

\begin{equation}
\begin{aligned}
g_T(c) &= 1-\frac{L_{\mathcal{T}}(c)}{L_{\mathcal{T}}(c_0)}, &
g_D(c) &= 1-\frac{D(c)}{D(c_0)},\\
g_E(c) &= 1-\frac{E(c)}{E(c_0)}, &
g_V(c) &= \frac{V(c)}{V(c_0)}-1.
\end{aligned}
\label{eq:pareto-gains}
\end{equation}

Candidates must preserve complete demand, throughput, teleports, and the global
and neighbor bounds in Equation~\ref{eq:top10-constrained-selection}.  A
candidate is Pareto-optimal when no other feasible candidate is at least as
good in all four gains and strictly better in one.  The baseline remains in
every pool.  This audit requires no additional LLM calls or SUMO runs and does
not alter the selected plan: it reports whether the plan chosen by the primary
Top-10 delay objective lies on the measured Pareto front.  Consequently, no
manually weighted preference profiles are introduced into the optimizer.

\section{Experimental Evaluation}
\label{sec:experiments}

We first compare the LLM-feedback optimizer with grounded random search and a
deterministic heuristic under the same microscopic-simulation budget.  We then
reuse the measured LLM candidate pools for an offline Pareto dominance audit.
All aggregates include no-operation outcomes as zero improvement and therefore
evaluate the complete guarded optimizer rather than only its successful
interventions.

\subsection{Experimental Setup}
\label{sec:experimental-setup}

We use five urban road networks derived from OpenStreetMap (OSM) data
(\copyright{} OpenStreetMap contributors)~\cite{openstreetmap} and three demand
seeds.  Each frozen demand file was generated with
\texttt{randomTrips.py}~1.24.0 and contains 2,400 passenger trips departing over
one hour.  Baselines and interventions are evaluated with Eclipse SUMO~1.27.1.
All feedback-refinement calls use the \texttt{gpt-5.6-luna} model.  The demand
is synthetic and is not calibrated to observed counts; the experiments
therefore test optimization behavior across network structures rather than
real-world traffic fidelity.  Table~\ref{tab:scenario-summary}
summarizes the scenarios and Table~\ref{tab:reproducibility-settings} summarizes the model, search-budget,
candidate-space, SUMO, and pairing settings used in the main comparison.

\begin{table}[!ht]
  \caption{Equal-demand SUMO scenarios. Delay is baseline mean vehicle time
  loss, reported as mean $\pm$ sample standard deviation across three seeds.}
  \label{tab:scenario-summary}
  \begin{center}
    \begin{tabular}{lrrrrr}
      \hline
      City & Edges & Junctions & TLS & Trips & Delay (s) \\
      \hline
      Albany      &  6,193 & 4,309 & 107 & 2,400 & $112.80 \pm 0.26$ \\
      Troy        &  6,107 & 4,355 &  57 & 2,400 & $ 99.79 \pm 0.84$ \\
      Ann Arbor   & 10,151 & 7,522 &  81 & 2,400 & $138.58 \pm 3.13$ \\
      Los Angeles &  6,969 & 4,954 & 172 & 2,400 & $175.76 \pm 1.26$ \\
      Manhattan   &  5,993 & 4,065 & 256 & 2,400 & $245.31 \pm 13.03$ \\
      \hline
    \end{tabular}
  \end{center}
\end{table}

\begin{table}[!ht]
  \caption{Reproducibility settings for the main comparison.}
  \label{tab:reproducibility-settings}
  \begin{center}
    \begin{tabular}{p{0.24\linewidth}p{0.67\linewidth}}
      \hline
      Component & Setting \\
      \hline
      Network source & Five urban road networks derived from OpenStreetMap
      data (\copyright{} OpenStreetMap contributors). \\
      LLM & OpenAI Chat Completions API; \texttt{gpt-5.6-luna}; model-default
      sampling; JSON response mode; 500 output tokens; 60-s timeout and at most
      two retries after the initial request, including one structured repair
      attempt for an invalid catalog selection. \\
      Search budget & At most seven candidate SUMO evaluations per case. LLM
      search uses three grounded initial proposals, up to two feedback
      mutations, and up to two compatible bundles; each baseline uses five
      proposals and up to two bundles. \\
      Candidate space & Signal timing, five-edge corridor speed control,
      single-edge rerouting, and validated topology actions. Numeric choices
      are discrete, and Top-10 ranking excludes edge fragments shorter than
      10~m. \\
      Signal timing & Fixed-time cycle and green-split plans derived using
      Webster's method; original phase states and yellow/all-red transition
      intervals are preserved. \\
      SUMO & Version 1.27.1; 1-s step; complete execution until all vehicles
      arrive; trip-level statistics and emissions plus 60-s edge measurements. \\
      Pairing & For each city and seed, all methods use the same demand, OD
      records, baseline diagnosis, objective, safety constraints, and SUMO
      seed. Only intervention-targeted artifacts are changed. \\
      \hline
    \end{tabular}
  \end{center}
\end{table}

Final selection minimizes time loss on the baseline-frozen Top-10 edges subject
to Equation~\ref{eq:top10-constrained-selection}.  A selected plan must preserve
demand and OD pairs, must not reduce throughput or increase teleports, and may
increase global mean delay and one-hop neighboring time loss by at most 1\% and
5\%, respectively.  The unmodified baseline is always an explicit candidate.

\subsection{Budget-Matched Proposal-Source Comparison}
\label{sec:proposal-source-comparison}

The \emph{LLM feedback} condition begins with one validator-confirmed random
proposal per intervention family.  After SUMO evaluation, the model selects a
single mutation from a finite legal catalog for each retained beam parent.  The
\emph{grounded random} baseline samples supported actions, diagnosed edge and
TLS identifiers, speed factors in $\{0.7,0.8,0.9\}$, and rerouting fractions in
$\{0.25,0.5,0.75,1.0\}$.  The \emph{deterministic heuristic} targets the
highest-ranked bottleneck for corridor and routing actions and the nearest TLS
in the diagnosed subgraph.  The downstream validator, bundle search, simulator,
and selector are shared by all three methods.

In addition to final Top-10 and global-delay reductions, we report proposal
validity and optimization success, defined as selecting a feasible plan with at
least 1\% Top-10 reduction.  Candidate-level simulation efficiency is
\begin{equation}
  \eta_{\mathrm{sim}}=
  \frac{N_{\mathrm{feasible},\,\Delta_{\mathrm{Top10}}\geq1\%}}
       {N_{\mathrm{SUMO}}}.
  \label{eq:simulation-efficiency}
\end{equation}

\begin{table*}[!ht]
  \caption{Budget-matched proposal-source comparison over five networks and three demand seeds. Success denotes a feasible selected plan with at least 1\% Top-10 reduction.}
  \label{tab:proposal-source-comparison}
  \begin{center}
    \small
    \setlength{\tabcolsep}{4pt}
    \resizebox{\textwidth}{!}{%
    \begin{tabular}{lrrrrrr}
      \hline
      Method & Valid (\%) & Success (\%) & Best nets. & Sim. eff. (\%) & Top-10 (\%) & Global (\%) \\
      \hline
      LLM feedback & \textbf{100.0} & \textbf{86.7} & \textbf{3/5} & 35.1 & 9.18 & 2.78 \\
      Grounded random & \textbf{100.0} & 73.3 & 1/5 & 34.0 & 9.00 & \textbf{3.03} \\
      Heuristic & 97.3 & 80.0 & 1/5 & \textbf{39.0} & \textbf{9.60} & 2.69 \\
      \hline
    \end{tabular}
    }
  \end{center}
\end{table*}

\begin{figure*}[!ht]
  \centering
  \includegraphics[width=0.96\textwidth]{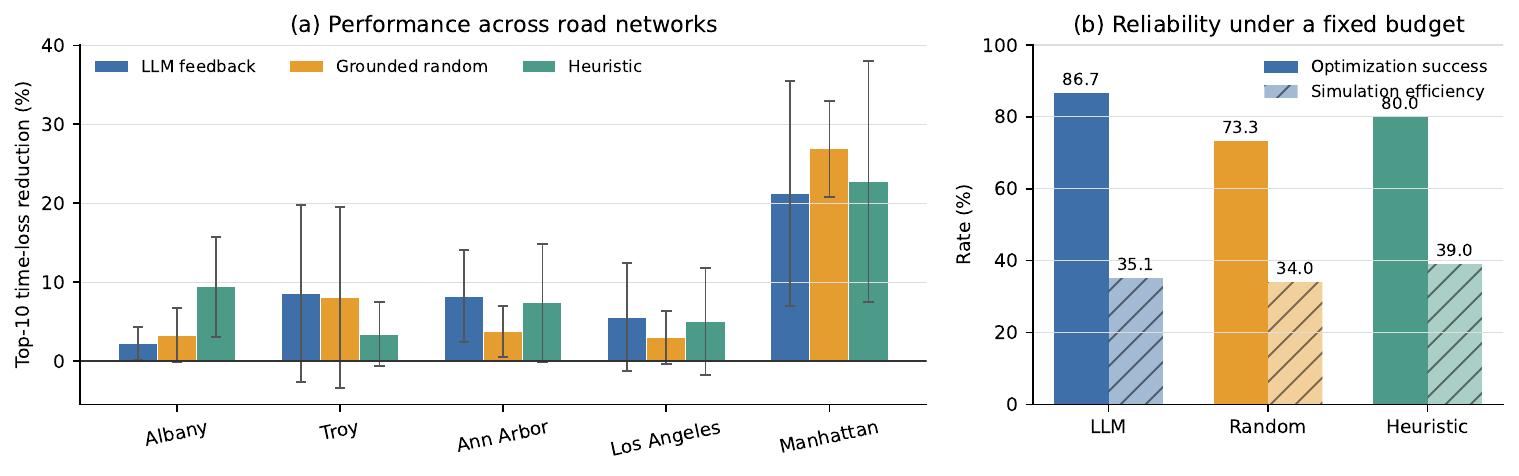}
  \caption{Proposal-source comparison over three demand seeds. Bars in (a)
  show mean Top-10 reduction and sample-standard-deviation error bars. Panel
  (b) reports case-level optimization success and candidate-level simulation
  efficiency under the common seven-evaluation cap.}
  \label{fig:proposal-source-comparison}
\end{figure*}

Table~\ref{tab:proposal-source-comparison} compares the three proposal sources
under the same simulation budget. Mean Top-10
reductions were $9.18\%\pm10.18\%$, $9.00\%\pm10.88\%$, and
$9.60\%\pm10.32\%$ for LLM feedback, grounded random, and the heuristic,
respectively.  Across the 15 city--seed cases, LLM feedback attained the highest optimization
success rate (86.7\%) and the lowest no-op rate (13.3\%).  It also achieved the
best seed-averaged Top-10 result in Troy, Ann Arbor, and Los Angeles, whereas
the heuristic led in Albany and grounded random led in Manhattan.  Their global-delay reductions were 2.78\%, 3.03\%, and 2.69\%.
The LLM's simulation efficiency was 35.1\%, between grounded random (34.0\%)
and the heuristic (39.0\%).  Six final LLM plans were feedback-refined singles;
together with four selected bundles, these results show that measured outcomes
changed the search even though the aggregate advantage was not significant.
Paired Wilcoxon tests on
Top-10 reduction gave $p=0.407$ against random and $p=0.639$ against the
heuristic.  Thus, the evidence supports broader cross-network success rather
than universal superiority over specialized heuristics.

Performance also varies with the demand realization.  Averaged over the five
networks, LLM feedback led seed 42 (12.38\% Top-10 reduction) and seed 44
(10.05\%), whereas the heuristic led seed 43 (9.03\%).  At the paired
city--seed level, LLM feedback beat, tied, and lost to grounded random in six,
six, and three cases, respectively.  It beat the heuristic in six cases and
lost in nine.  The high cross-case standard deviations in
Figure~\ref{fig:proposal-source-comparison} therefore reflect both demand-seed
sensitivity and genuine differences between network structures, particularly
the large attainable improvements in Manhattan.

Table~\ref{tab:five-city-llm-results} reports the proposed optimizer separately
for each network.  Ann Arbor, Los Angeles, and Manhattan accepted a feasible
improving plan for all three demand seeds; Albany and Troy each retained the
baseline once.  Manhattan provides the largest mean reduction and the highest
candidate yield, while the smaller Albany improvements illustrate that the
baseline guard does not force an intervention when the legal search space
offers only marginal or unsafe changes.

\begin{table}[!ht]
  \caption{Five-network performance of the LLM-feedback optimizer over three
  demand seeds. Reductions are mean $\pm$ sample standard deviation in
  percentage points and include no-op outcomes.}
  \label{tab:five-city-llm-results}
  \centering
  \small
  \setlength{\tabcolsep}{3pt}
  \begin{tabular}{lrrrr}
    \hline
    Network & Top-10 (\%) & Global delay (\%) & Sim. eff. (\%) & Success \\
    \hline
    Albany & 2.25 $\pm$ 2.11 & 0.17 $\pm$ 0.52 & 21.1 & 2/3 \\
    Troy & 8.55 $\pm$ 11.17 & 0.75 $\pm$ 1.27 & 28.6 & 2/3 \\
    Ann Arbor & 8.25 $\pm$ 5.83 & 1.75 $\pm$ 0.46 & 30.0 & 3/3 \\
    Los Angeles & 5.58 $\pm$ 6.84 & 0.51 $\pm$ 1.28 & 31.6 & 3/3 \\
    Manhattan & 21.27 $\pm$ 14.23 & 10.71 $\pm$ 7.55 & 66.7 & 3/3 \\
    \hline
    Overall & 9.18 $\pm$ 10.18 & 2.78 $\pm$ 5.08 & 35.1 & 13/15 \\
    \hline
  \end{tabular}
\end{table}

\subsection{Search Behavior and Intervention Composition}
\label{sec:search-behavior}

\begin{figure*}[!ht]
  \centering
  \includegraphics[width=0.92\textwidth]{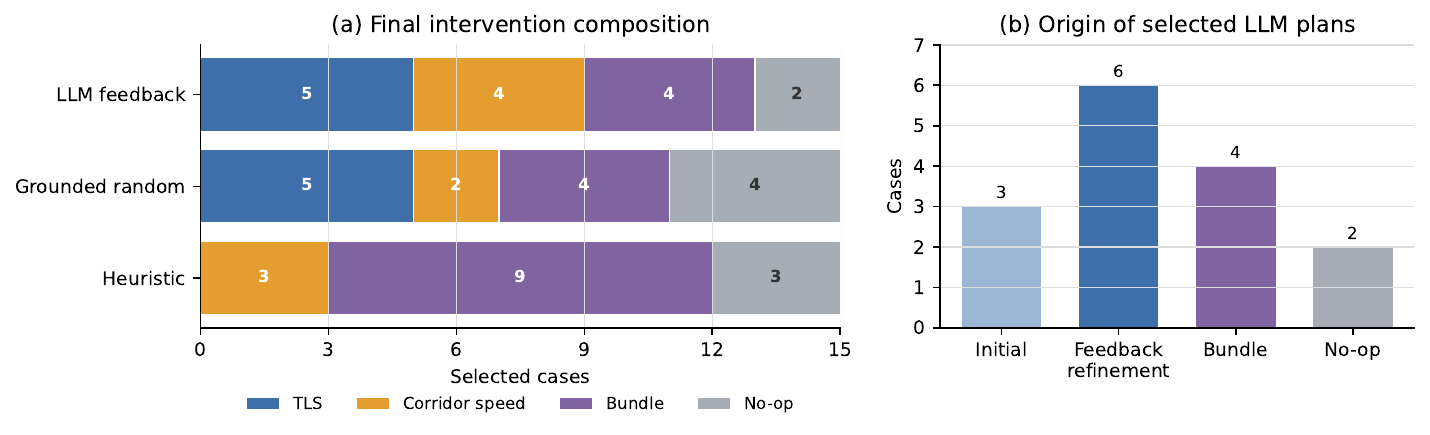}
  \caption{Search outcomes over 15 city--seed cases. Panel (a) reports the
  intervention family selected by each proposal source. Panel (b) separates
  initial, feedback-refined, bundled, and no-op outcomes for the LLM search.}
  \label{fig:proposal-source-actions}
\end{figure*}

Figure~\ref{fig:proposal-source-actions} examines how each method reaches its
final decision.  LLM feedback selected five TLS plans, four corridor-speed
plans, four bundles, and two no-ops.  Grounded random selected the same number
of TLS plans and bundles but retained the baseline in four cases, consistent
with its lower optimization-success rate.  The deterministic heuristic was
more concentrated: nine of its 12 applied plans were bundles and the remaining
three used corridor-speed control.  Routing and topology candidates were
evaluated, but none won the final constrained selection because their local
benefits did not compensate for global-delay, spillover, or demand-preservation
failures.

Only three of the 13 applied LLM plans were unmodified initial singles.  Six
were produced by simulation-feedback refinement and four were selected by
bundle search.  Thus, ten applied decisions required a second-stage search
operation rather than merely accepting the grounded random initialization.
This distinction is important because the aggregate comparison otherwise
understates how often measured simulator feedback changes the executable plan.

\subsection{Pareto Dominance Audit}
\label{sec:pareto-results}

We audit the same 15 LLM candidate pools without additional model calls or SUMO
runs.  Dominance is evaluated jointly over Top-10 delay, global delay,
$\mathrm{CO}_2$, and mean speed after applying the operational constraints.
The resulting fronts contain 1.53 candidates on average (range 1--4), and 11 of
the 15 plans selected by the primary Top-10 rule are nondominated.  The four
dominated selections show that optimizing bottleneck delay can sacrifice a
secondary metric, while the generally small fronts indicate limited
multi-objective diversity in the current delay-oriented candidate pools.  We
therefore report Pareto status as a diagnostic rather than introducing
manually weighted profiles or changing the optimizer's primary objective.

\section{Conclusions}
\label{sec:conclusion}

This study presented a simulation-grounded multi-agent LLM framework for
traffic intervention optimization in SUMO.  The framework separates grounded
initialization, feedback-guided local refinement, and final decision making.  A
seeded sampler constructs initial signal-control, corridor-speed, and routing
actions, after which role-specific LLM agents use measured SUMO outcomes to
select one-parameter refinements from validator-confirmed catalogs.
Deterministic utilities materialize the actions, and every valid single action
and compatible pairwise bundle is evaluated through a complete simulation.
The returned plan is the best evaluated candidate satisfying constraints on
global delay, neighboring-road spillover, throughput, demand preservation, and
teleport events; otherwise, the unmodified scenario is retained.

Across five road networks and three demand seeds, the guarded optimizer reduced
Top-10 bottleneck time loss by $9.18\%\pm10.18\%$ and network-wide delay by
2.78\%.  It selected a feasible improving plan in 86.7\% of cases, compared
with 73.3\% for grounded random search and 80.0\% for the deterministic
heuristic under the same seven-simulation cap.  Only three of the 13 applied
LLM plans were unmodified initial actions; six were feedback-refined singles
and four were bundles, showing that the second-stage search frequently changed
the executable decision.  However, paired Wilcoxon tests found no statistically
significant Top-10 advantage over grounded random search ($p=0.407$) or the
deterministic heuristic ($p=0.639$); the heuristic also produced a slightly
larger mean Top-10 reduction.  The evidence therefore supports improved
feasible-plan coverage and competitive search performance rather than universal
superiority of LLM-guided refinement.

The current findings are limited by synthetic demand, three demand seeds, one
LLM model and delay-oriented feedback prompt, and a small simulation budget.
The comparison also does not isolate LLM mutation selection from the surrounding
beam-search structure using identical-parent random and rule-based selectors,
and it does not include an established traffic-control or simulation-optimization
baseline.  Future work should address these limitations using calibrated or
observed demand, additional LLM and demand realizations, matched mutation-selector
ablations, and conventional traffic-control benchmarks.  Larger bundles,
coordinated controllers, and time-varying policies also remain open extensions.
Overall, the results position LLMs as constrained, feedback-guided local search
operators rather than autonomous traffic controllers: executable tools,
microscopic simulation, and explicit safety constraints retain authority over
the final intervention.

\section*{AUTHOR CONTRIBUTIONS}
The authors confirm their contributions to the paper as follows: Shuyang Li:
conceptualization, methodology, software development, investigation, formal
analysis, visualization, and writing---original draft. Ruimin Ke: supervision
and writing---review and editing. Both authors reviewed and approved the final
version of the manuscript.

\section*{GENERATIVE AI DISCLOSURE}
The authors reviewed and verified the
resulting text and take full responsibility for the manuscript's content. The
use of \texttt{gpt-5.6-luna} as an experimental component of SimTIO is reported
separately in the Methodology section.




\newpage
\bibliographystyle{chicago}
\bibliography{trb_template}
\end{document}